\documentclass[10pt,conference]{IEEEtran}
\IEEEoverridecommandlockouts
\usepackage[utf8]{inputenc}
\usepackage{graphicx} 
\usepackage{booktabs} 
\usepackage{amsmath}
\usepackage{bm}
\usepackage{gensymb}
\usepackage{physics}
\usepackage{mathtools, nccmath}
\usepackage{algorithm}
\usepackage{algpseudocode}
\usepackage{setspace}
\usepackage{amsfonts}
\usepackage{amssymb}
\usepackage{bbm}
\usepackage{cite}
\usepackage{lipsum}
\usepackage{authblk}
\usepackage{adjustbox}
\usepackage{comment}
\usepackage{hyperref}

\title{A Fully Asynchronous \\ Unsourced Random Access Scheme}
\author[*]{Mert Ozates\thanks{Mert Ozates' work is supported by European Commission’s Horizon Europe, Smart Networks and Services Joint Undertaking, research and innovation program under grant agreement number 101139282, 6G-SENSES project. Gianluigi Liva's work is supported by the Federal Ministry of Education and Research of Germany in the programme of “Souverän. Digital. Vernetzt” joint Project 6G-RIC, project identification number: 16KISK026. Mohammad Kazemi’s work was funded by UK Research and Innovation (UKRI) under the UK government’s Horizon Europe funding guarantee [grant number 101103430].}}
\author[**]{Mohammad Kazemi}
\author[***]{Gianluigi Liva}
\author[**]{Deniz Gündüz}
\affil[*]{IHP - Leibniz Institute for High Performance Microelectronics, 15236 Frankfurt (Oder), Germany \protect\\ Email: oezates@ihp-microelectronics.com}
\affil[**]{Department of Electrical and Electronic Engineering, Imperial College London  \protect\\ Email:\{mohammad.kazemi, d.gunduz\}@imperial.ac.uk}
\affil[***]{Institute for Communications and Navigation, German Aerospace Center (DLR), Wessling, Germany \protect\\Email: gianluigi.liva@dlr.de  }
\date{}

\begin{document}

\maketitle
\thispagestyle{empty}
\pagestyle{empty} 

\begin{abstract}
    We investigate fully asynchronous unsourced random access (URA), and propose a high-performing scheme that employs on-off division multiple access (ODMA). In this scheme, active users distribute their data over the transmit block based on a sparse transmission pattern without any limitations on the starting time. At the receiver side, we adopt a double sliding-window decoding approach, utilizing a smaller inner decoding window of two block lengths within a larger outer window to enhance the interference cancellation process. Within the inner window, the receiver iteratively applies preamble-free joint starting time and pattern detection, single-user decoding, and successive interference cancellation operations. A notable feature of the proposed scheme is its elimination of the need for a preamble for starting time detection; this is achieved using ODMA transmission patterns. Numerical results demonstrate that the proposed asynchronous URA scheme outperforms existing alternatives.
\end{abstract}


\section{Introduction}

 In some applications of beyond 5G and 6G communication systems, there may be millions of sporadically active devices, making any form of coordination impractical. Specifically, it is not feasible to schedule users or assign dedicated resources due to latency and feedback overhead. To tackle these challenges and facilitate communication among a vast number of devices, Polyanskiy introduced the unsourced random access (URA) paradigm in \cite{polyanskiy}. In \cite{polyanskiy}, it is proposed that devices use a common codebook for transmission, meaning there is no user identity, allowing system to operate independently of the total number of devices. Decoding is performed to extract a list of transmitted messages up to a permutation, with a per-user probability of error (PUPE)  as the main performance criterion.


Conventional random access schemes like ALOHA become energy inefficient as the number of users increases \cite{polyanskiy}, which makes developing low-complexity and energy-efficient transmission schemes for URA an important problem. This can be addressed by separating users in the time domain by slotting the transmission frame \cite{vem}, utilizing coded compressed sensing \cite{comp2}, random spreading \cite{pradhan,schiavone}, or on-off division multiple access (ODMA) \cite{odma1,odma2}, where a small fraction of the transmission frame is allocated to each user based on a transmission pattern. While the authors of \cite{vem,comp2,pradhan,schiavone,odma1,odma2} study URA over a Gaussian multiple access channel (MAC), the fading setup is investigated with a single-antenna receiver in \cite{fading1}, and with a base station (BS) equipped with a massive number of antennas in \cite{fasura2,orth,twc,odma4}.

The aforementioned works assume that the active users transmit in a frame-aligned manner, meaning the transmission is synchronous; hence, they must listen for a beacon from the receiver. In some systems, nevertheless, it is more practical to remove the need for time synchronization to a common reference. For example, this is the case where one wants to operate nodes with limited (or no) reception capabilities. Asynchronous random access protocols are particularly appealing in long-range wireless terrestrial \cite{lora} and satellite-based \cite{scalise} IoT/mMTC systems, where ensuring tight synchronization of user terminals to a common time reference comes at a significant cost, in terms of both energy and protocol overhead. In the context of satellite communications, powerful, asynchronous massive random access schemes were introduced, for example, in \cite{herrero, clazzer}.





Asynchronous access is particularly desirable for URA, due to the huge number of devices in the system, many of which are likely to be cheap and low-power. However, there have been limited efforts on asynchronous URA to date. In \cite{kowshik}, the authors consider fading MAC and use OFDM to mitigate the effects of synchronization errors, assuming that the maximum error is less than the cyclic prefix length. Asynchronous URA is also addressed in 
\cite{mimoasync1,mimoasync2} with the help of OFDM, similarly to 
\cite{kowshik}, utilizing a massive MIMO receiver. While limited asynchronism is presumed in 
\cite{kowshik,mimoasync1,mimoasync2}, a fully asynchronous setup is explored in 
\cite{karami}, where there are no restrictions on transmission start times, allowing active users to send their packets on-demand. In other words, they can begin their transmission at any random time. Several works showed that asynchronism can improve performance in the (unsourced) MAC setting. In \cite{hou}, it is shown that asynchronism can help a successive interference cancellation (SIC) receiver to separate user transmissions in a coordinated MAC scenario. Variations on coded slotted ALOHA \cite{sandgren,clazzer} showed how asynchronism can be effectively leveraged to improve the throughput of synchronous coded random access protocols. Furthermore, the authors of  \cite{fengler} demonstrate how the sum-rate of a two-user binary adder unsourced MAC can be improved, with binary linear block codes, by introducing a sufficient amount of asynchronisms between user transmissions.




In this paper, we examine the fully asynchronous setup and propose an energy-efficient solution. We assume that active users encode their messages using a polar code and transmit the polar codewords by employing ODMA. Specifically, the codeword symbols of each user are placed in a small fraction of the time instances within their transmitted packet, following a transmission pattern, while the rest of the packet remains idle. At the receiver, we utilize a sliding window decoding approach that employs one inner and one outer window. Within the inner sliding window, iterative decoding is applied, where the starting times and transmission patterns of the active users are jointly estimated in the first step using the received signal power of the patterns, as our proposed scheme does not include a preamble. Subsequently, a single-user polar decoder is employed for each user, followed by SIC. Numerical examples show that our proposed scheme outperforms the one in \cite{karami} by up to 5.5 dB when the number of packet arrivals per packet duration is less than or equal to 100.


The remainder of the paper is organized as follows: we present the system model in Section \ref{system} and the proposed scheme in Section \ref{proposed}. We provide a set of numerical results in Section \ref{results} and draw conclusions in Section \ref{conclusion}.


\section{System Model} \label{system}

We consider an asynchronous URA setup where active users transmit $B$ bits of information over a real-valued Gaussian MAC to a common BS equipped with a single antenna. We assume that the transmission is frame-asynchronous (but symbol-synchronous\footnote{The assumption of symbol synchronism is used only to simplify the analysis, and it can be relaxed without impacting the outcome of the analysis.}) and there is no limitation on the starting times of user transmissions. Specifically, a user sends its packet whenever it has data to transmit. In this model, the received signal at the $j$-th time instance can be expressed as

\begin{equation}
    \mathbf{y}_j = \sum\limits_{i \in \mathcal{K}_j} \mathbf{x}_i(j-\delta_i) + \mathbf{z}_j,
\end{equation}

\noindent where $\delta_i$ denotes the transmission starting time of the $i$-th user, $ \mathcal{K}_j $ is the set of active users interfering at the $j$-th instance, and $\mathbf{z}_j$ is the additive white Gaussian noise (AWGN) with zero-mean and variance $\sigma^2$. The packet of the $i$-th user is denoted by $\mathbf{x}_i$, with a packet length of $n$. It consists of $n_d$ codeword symbols (active indices), while the remaining $n - n_d$ indices are idle, and it satisfies the power constraint $\norm {\mathbf{x}_i}^2 \leq n_dP$.


At the receiver, a list of transmitted messages $\mathcal{L}$ is extracted up to a permutation. We assume, w.l.o.g., there is a transmission that begins at time $t = 0$. We compute the PUPE by observing the channel output over a long time interval $[0, T + n]$, where $T \gg n$. It can be written as


\begin{equation}
P_e =\frac{1}{K_{a,T}} \sum\limits_{i=1}^{K_{a,T}} \Pr(\mathbf{m}_i \notin \mathcal{L}),
\end{equation}


\noindent where $K_{a,T}$ is the number of active users in $[0, T]$, i.e., the number of users initiating a transmission in $[0, T]$, and $\mathbf{m}_i$ denotes the message from the $i$-th user. It is important to note that the average number of active user arrivals per packet duration, i.e., the normalized active user load is indicated by $K_a$.\footnote{This is equivalent to the definition of channel load in the ALOHA literature.} The energy-per-bit of the system can be calculated as


\begin{equation}
    \frac{E_b}{N_0} =  \frac{n_dP}{2B\sigma^2}.
\end{equation}

The goal is to minimize the required energy-per-bit while meeting a target PUPE of $\epsilon$.

\begin{figure}
    \centering
    \hspace*{-5mm}
     \includegraphics[scale = 0.35]{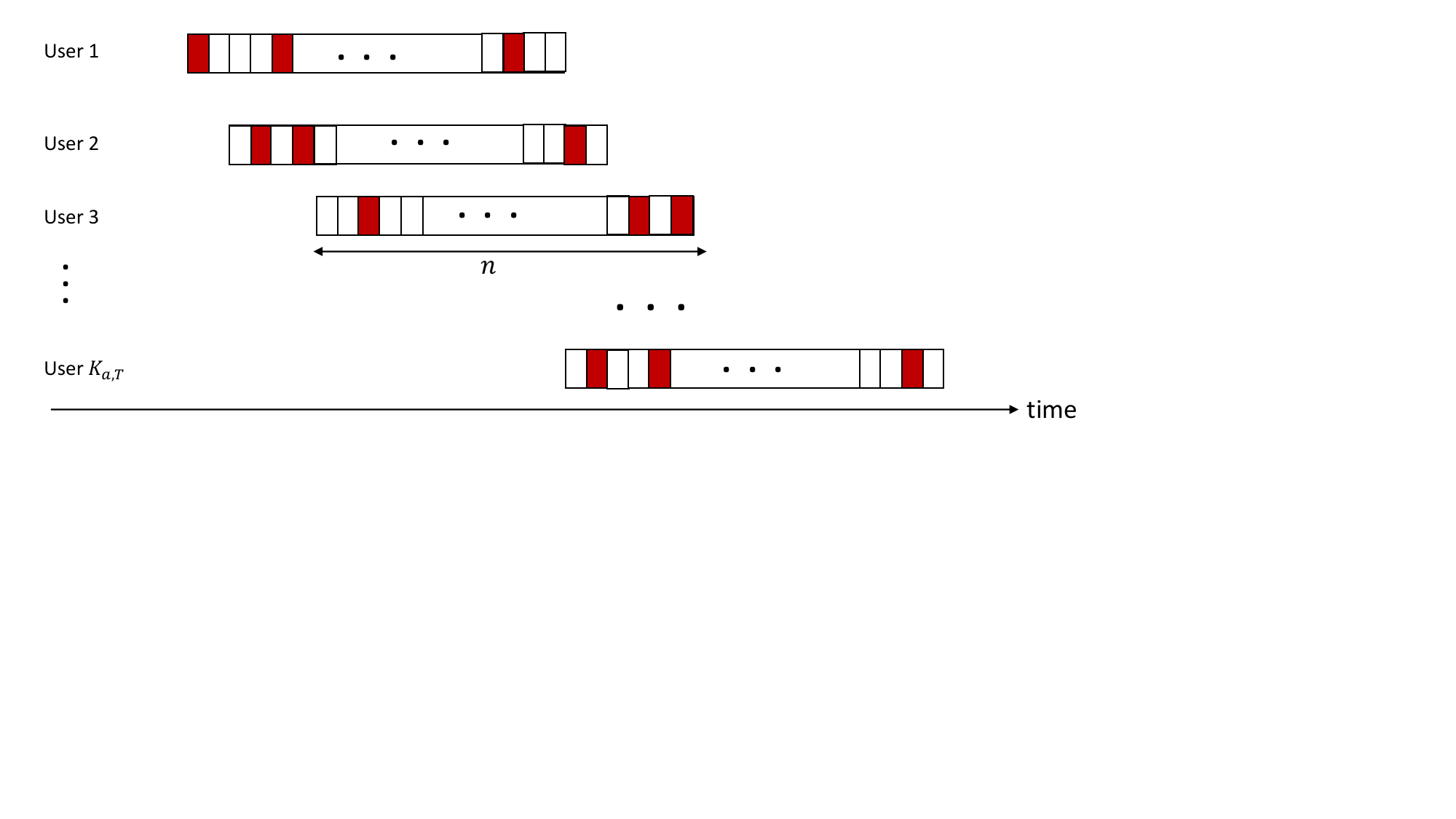}
     \vspace*{-40mm}
    \caption{The fully asynchronous ODMA-based transmission structure. Colored boxes illustrate the utilized channel symbols by each active user.}
    \label{encoder}
\end{figure}

\section{Proposed Scheme} \label{proposed}

\subsection{Encoding}

\begin{figure*}
    \centering
     \includegraphics[scale = 0.43]{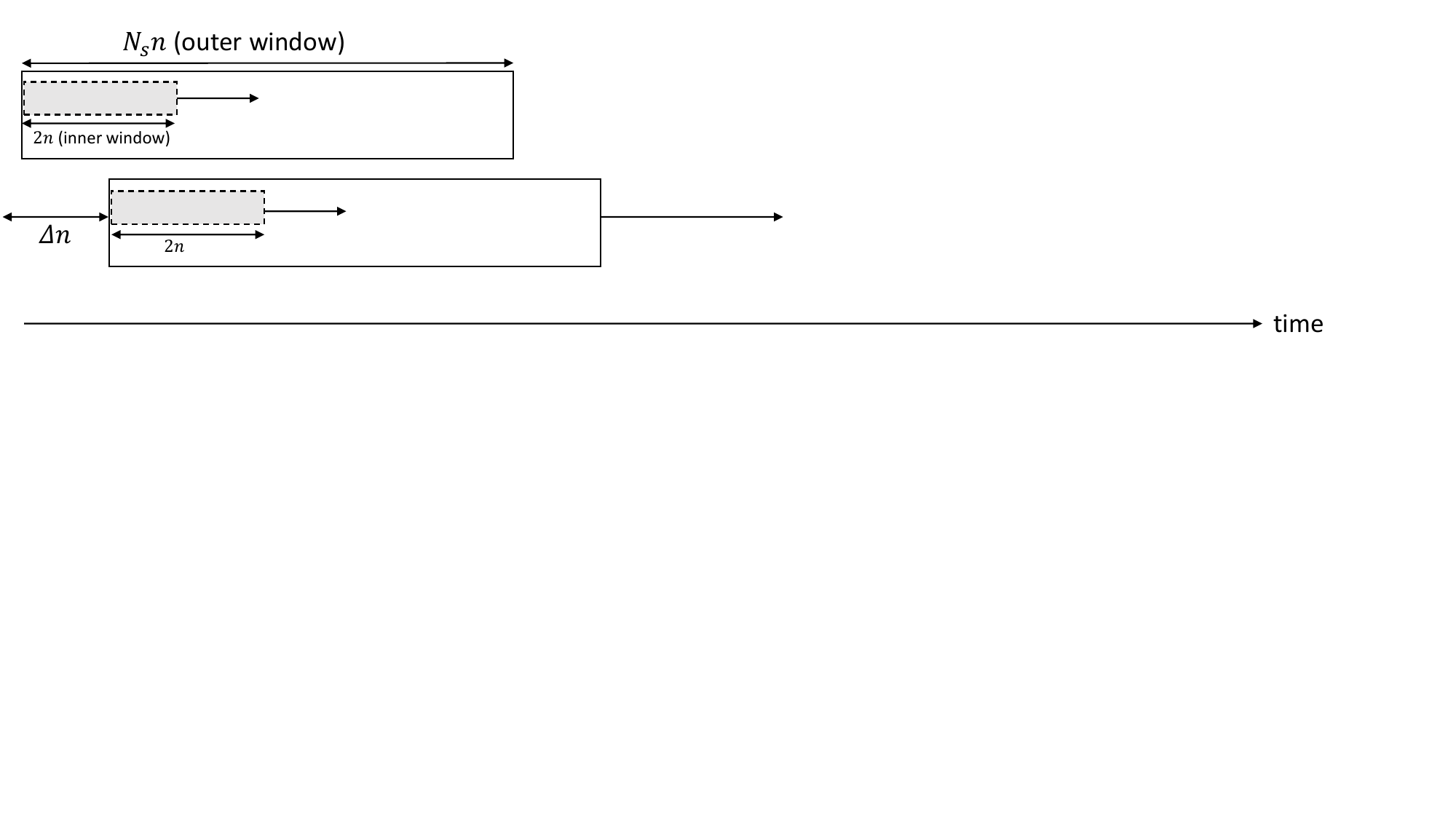}
    \vspace*{-53mm}
    \caption{Decoding process with two sliding windows.}
    \label{figslide}
\end{figure*}

In our proposed scheme, each user's message sequence is divided into two parts of lengths $B_p$ and $B_c = B-B_p$. The second part is appended with $r$ cyclic redundancy check (CRC) bits, encoded using a polar code with block length $n_c$ and modulated by binary phase shift keying (BPSK). The encoded and modulated sequence is then transmitted via ODMA; its elements occupy $n_d$ of $n$ channel uses ($n_d = n_c$ for BPSK) after the transmission starting time, following a transmission pattern determined by the first $B_p$ bits. The transmission patterns are chosen randomly and independently from a common pattern matrix $\mathbf{P} \in \{0,1\} ^{n \times M_p}$, where $M_p = 2^{B_p}$. Each column of $\mathbf{P}$ contains $n_d$ non-zero elements that determine the placement of transmitted symbols, while the remaining elements represent the idle indices. The transmitted signals of the active users are illustrated in Fig. \ref{encoder}, depicting the ODMA concept and fully asynchronous transmission.


\subsection{Decoding}

At the receiver side, we employ a sliding window-based decoding approach to recover the transmitted messages. In the proposed decoder, two nested sliding windows are utilized: the first is the outer sliding window of $N_s$ packet lengths, and the second is the inner sliding window of two packet lengths, which is taken to be the minimum (assuming that the window length is an integer multiple of the packet length for simplicity) to increase the detection probability in pattern estimation, to be explained later. The decoding operations of joint starting time and active pattern estimation, single-user decoding, and SIC occur in each iteration of the inner sliding window operation, aiming to decode the active users starting their transmissions between the first and $n$-th time instances of the inner window. Once the iterations are completed, the inner window moves by one packet length, and the same operations are repeated. When the inner window reaches the end of the outer window, one outer decoding iteration is completed. Then, the inner window returns to the start of the outer window, and the same procedure is repeated. When the outer iterations are completed, the outer window shifts by $\Delta n$ (i.e., $\Delta$ packet lengths), and the decoding operation in the outer window is applied. The nested operation of the sliding windows is illustrated in Fig. \ref{figslide}. In the rest of the subsection, we present the decoding operations in the inner sliding window in detail.


\subsubsection{Joint Starting Time and Pattern Estimation}

To detect the transmitted messages, we first jointly estimate the starting times and transmission patterns of the active users using only the received signal, that is, without any preamble. To estimate these parameters, we test each pattern at each time instance $b = 1, 2, \dots, n$, by calculating the received power at their active indices as follows.

\begin{equation}
    e_{i,b} = \norm{\mathbf{y}_b \odot  \mathbf{p}_{i}}_1, \quad i=1,2 \dots M_s,
    \label{eqpattern}
\end{equation}



\noindent where $\mathbf{y}_b$ represents the length-$n$ segment of the received signal in the inner sliding window $\mathbf{y}_s$, beginning at time instance $b$, $\odot$ denotes elementwise multiplication, $\norm{.}_1$ denotes the $l1$-norm, and $\mathbf{p}_{i}$ denotes the $i$-th column of $\mathbf{P}$. To simplify the notation, we omit the iteration index here, and the equation is presented for the first iteration; however, the same operation is carried out with the residual signal in subsequent iterations. We assume that the transmission pattern with the maximum signal powers is the survivor pattern at that time instance. After repeating this procedure for each of the first $n$ time instances in the inner sliding window, we select the time indices of the $K_a + u$ survivor patterns with maximum power as the estimated starting time instances, where $u$ is a margin on the average active user load, accounting for the variation in the number of active users initiating their transmissions within each time block of length $n$.

Another (and the more conventional) way to estimate the starting times in an asynchronous setup is to append a preamble to all codewords, and use it for timing detection. For this purpose, one can calculate a simple correlation of the preamble with the received signal as

\begin{equation}
    c_{b} = \sum\limits_{i = 1}^{n_p} \mathbf{y}_b(i) \mathbf{a}(i),  \quad b=1,2 \dots n, 
    \label{eqtime}
\end{equation}

\noindent where $\mathbf{a}$ is the preamble, $\mathbf{a}(i)$ denotes its $i$-th element, $\mathbf{y}_b(i)$ is the $i$-th element of  $\mathbf{y}_b$, and $n_p$ is the preamble length. Then, $K_a + u$ points that maximize the correlation in (\ref{eqtime}) can be taken as the starting time estimates, and the transmission patterns are estimated using (\ref{eqpattern}) for each detected starting time. Note that the alternative scheme for fully asynchronous URA in 
\cite{karami} uses preambles as described above to estimate the transmission starting times. Removing the preamble offers several potential advantages: there is no energy overhead due to the preamble, and its potential interference is eliminated. However, the detection performance of the starting times may suffer, as it relies on the received signal power of the patterns, some of which might be obscured by interference. To mitigate this effect, we use an inner sliding window with a minimum length since our extensive simulations indicate that the performance of the proposed starting time and pattern detection algorithm declines when there are more options for the starting times.



\subsubsection{Single-user Decoding and SIC}

\begin{figure}
    \centering
    \includegraphics[scale = 0.45]{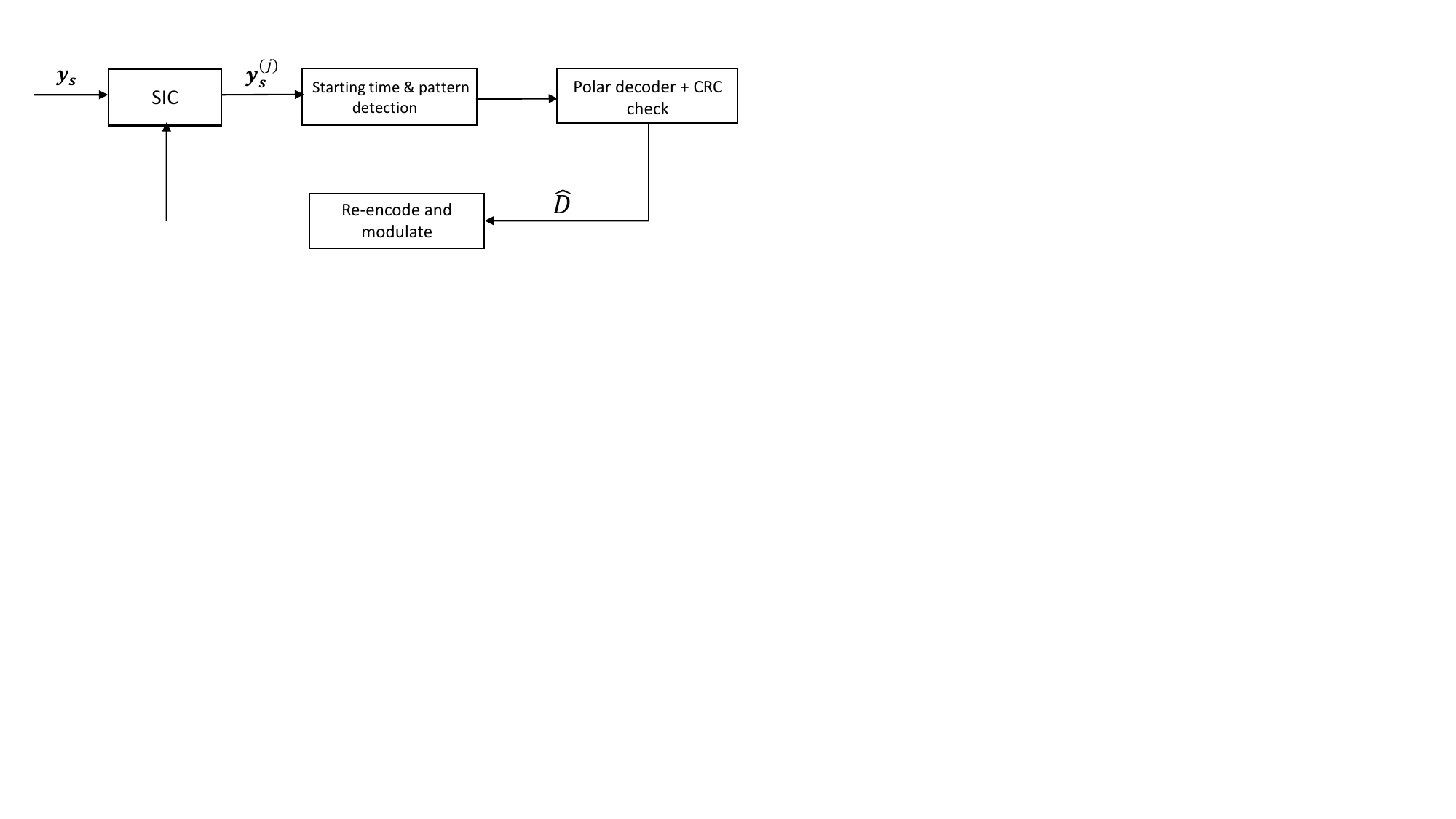}
     \vspace{-65mm}
    \caption{The decoding process within the inner sliding window.}
    \label{decoder}
\end{figure}

Once the starting times and transmission patterns are detected, for each pair of starting time and transmission pattern, the corresponding received signal points are collected, and the log-likelihood ratios (LLRs) for that user are extracted while the interference of other users is treated as Gaussian noise. The LLRs are input into a single-user polar decoder that employs successive cancellation list decoding (SCLD). If the CRC check is satisfied, the decoding is considered successful; the decoded message is added to the temporary output list $\mathcal{\hat{D}}$, and its effect is subtracted from the received signal after re-encoding and modulation. The decoding iterations continue until no new message can be successfully decoded in the current iteration, or until $n_{\text{max}}$ iterations are completed. Then, the inner sliding window is shifted by one packet length of $n$ channel symbols, and the same procedure is repeated until the outer window ends, (i.e., $N_s -1$ times) which can be regarded as one outer iteration. The decoding operation inside the inner sliding window is illustrated in Fig. \ref{decoder}. Moreover, two pseudo-codes for the decoding operations of the inner sliding window and the whole process are provided in Algorithm \ref{alg_inner} and Algorithm \ref{alg_outer}, respectively, where $n_{\text{out}}$ is the number of outer decoding iterations.






 \begin{algorithm}[t]
\caption{Decoding operations in the inner sliding window.}\label{alg_inner}
\begin{algorithmic}[1]

\State \textbf{Input}: $\mathbf{y}_s$,  $\mathbf{P}$, $u$, $n_{\text{max}}$
\State \textbf{Iterative decoding}:
\State Set $\mathcal{\hat{D}} = \emptyset$.
\For{$j=1,2, \ldots n_{\text{max}} $}
\State \textbf{Joint starting time and pattern estimation}:

\For{$k=1,2, \ldots n $}

\State $\mathbf{y}_b$ = $\mathbf{y}_s[k:k+n-1]$.

\For{$k=1,2, \ldots M_s $}
\State $ e_{i,b} = \norm{\mathbf{y}_b \odot  \mathbf{p}_{i}}_1 \quad i=1,2 \dots M_s, $
\EndFor
\EndFor
\State Set time indices of maximum $K_a + u$ survivor patterns as estimated starting times.
\State \textbf{Single-user decoding and SIC}:
\State Set $d = 0$.
\For{$i=1,2,\ldots, {K}_a + u$}
\State Extract LLRs by treating interference as noise.
\State Run polar decoder on LLRs $\rightarrow$ $\mathbf{\hat{m}}_i$.
\If {CRC check is successful:}
\State Add $\mathbf{\hat{m}}_i$ to $\mathcal{\hat{D}}$.
\State d = d + 1.
\State Apply SIC.
\EndIf
\EndFor
\If {$  d = 0$}
\State Terminate the algorithm.
\EndIf
\EndFor


\State $\textbf{Output}$: Intermediate list of the decoded messages $ \mathcal{\hat{D}} $


\end{algorithmic}
\end{algorithm}

\begin{algorithm}[t]
\caption{Decoding process of the proposed scheme.}\label{alg_outer}
\begin{algorithmic}[1]

\State \textbf{Input}: $\mathbf{y}$, $N_s$,  $n_{\text{out}}, T, \Delta$


\State $t' = 1$

\While{$t' < T-(N_s-1)n + 1$} (outer sliding window)

\State $ w_{\text{str}} = t'$, $w_{\text{end}} = t' + N_sn-1$
 
\State $\mathbf{y}_w$ = $\mathbf{y}[w_{\text{str}}:w_{\text{end}}]$.

\For{$i=1,2, \ldots ,n_{\text{out}} $} (outer iterations)
\For{$j=1,2, \ldots ,N_s-1 $} (inner sliding window)

\State $s_{\text{str}} = (j-1)n + 1$, $s_{\text{end}} = s_{\text{str}} + 2n-1$.
\State $\mathbf{y}_s = \mathbf{y}_w [s_{\text{str}}:s_{\text{end}}] $.
\State \textbf{Perform Algorithm \ref{alg_inner}}. 

\EndFor
\EndFor

\State $t' = t' + \Delta n$

\EndWhile

\State $\textbf{Output}$: Final list of the decoded messages $\mathcal{L}$

\end{algorithmic}
\end{algorithm}




\vspace*{-5mm}
\section{Numerical Results} \label{results}

In this section, we evaluate the performance of the proposed scheme. We set the packet length to $n = 10000$, the message length to $B = 100$ bits, and the target PUPE to $\epsilon = 0.05$, as in \cite{karami} for direct comparison. For transmission patterns, we use a randomly generated binary matrix of size $n \times M_p$ with a column weight of $n_d$, where $M_p = 16$ ($B_p = 4$). It is important to mention that one can utilize a small number of transmission patterns as the pattern collisions can lead to a decoding failure only when two users having the same pattern start their transmissions at the same time instance, which has a very low probability. We utilize 5G polar codes of length 512 or 256 depending on the scenario, set the CRC length to 16, the list size of SCLD to 32, $n_{\text{max}} = 50$, and $n_{\text{out}} = 10$. We define the outer sliding window length as $N_s = 5$ packet lengths and shift the outer window by one packet length after completing the outer iterations (i.e., $\Delta n = n$). Note that to enhance decoding performance, we apply power diversity among the users by dividing the transmission patterns into equally sized groups and adjusting the transmitted signal of each user based on its transmission pattern index while adhering to a total power constraint. 

\begin{figure} 
    \centering
    \includegraphics[width=1\linewidth]{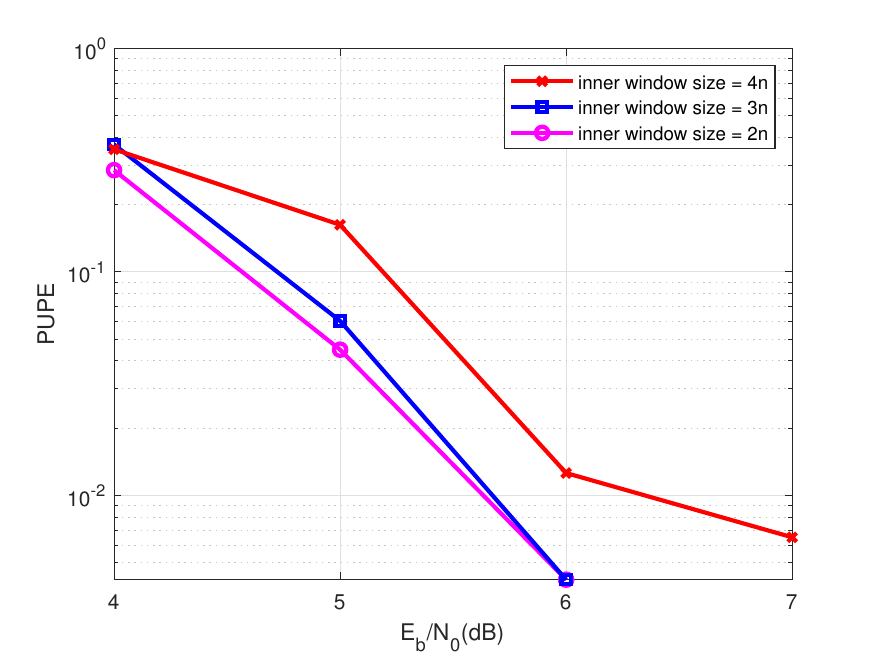}
    \caption{PUPE comparison for different inner sliding window lengths for $K_a = 75$ and outer window length $N_s = 5$.}
    \label{figinner}
\end{figure}

We first compare the performance of the proposed scheme with different inner sliding window sizes in Fig. \ref{figinner} for $K_a = 75$. The results in Fig. \ref{figinner} show that the inner window of length $2n$ has the best performance, which justifies our selection of taking its length as the minimum. The reason behind this behavior is that the performance of starting time and pattern detection degrades with the increasing inner window size, and PUPE is lower bounded by the detection error in this step. 

\begin{figure} 
    \centering
    \includegraphics[width=1\linewidth]{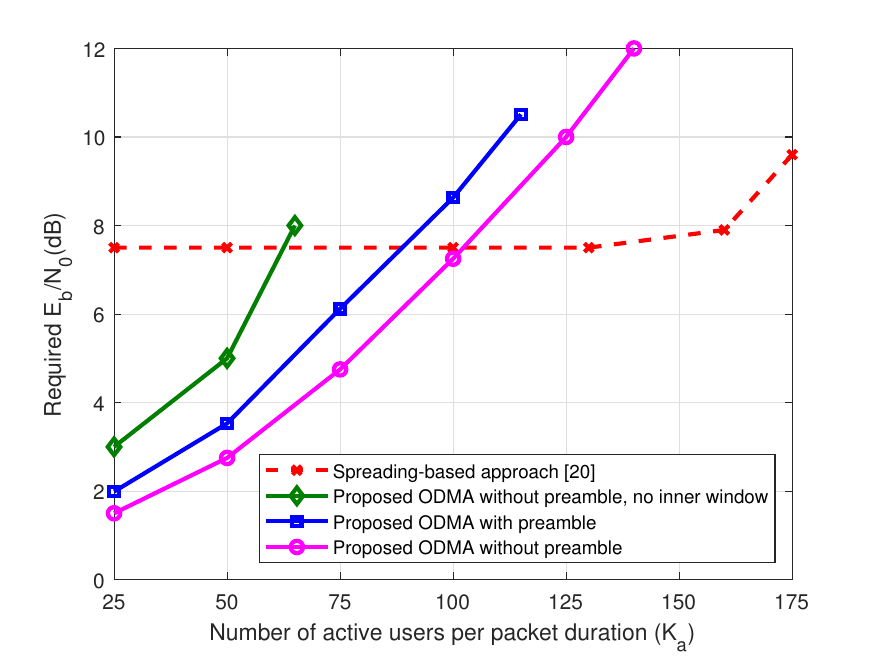}
    \caption{Comparison of the required $E_b/N_0$ versus the number of active users per packet duration  for $P_e \leq 0.05.$}
    \label{figeff}
\end{figure}

We then compare the performance of the proposed scheme with that of \cite{karami} in Fig. \ref{figeff}. The results in Fig. \ref{figeff} show that our proposed scheme outperforms the scheme in \cite{karami} for $K_a \leq 100$ by up to 5.5 dB; however, its performance is inferior to that of \cite{karami} for $K_a > 100$. We also implement our scheme with a preamble for estimation of the starting times to highlight the benefits of our proposed preamble-free scheme and show that our proposed scheme outperforms its preamble-based version by up to 1.5 dB. Fig. \ref{figeff} also illustrates the importance of the inner sliding window in the proposed scheme, as it offers much better performance compared to its single-window version.



\section{Conclusions} \label{conclusion}

We examined URA in a fully asynchronous setup where active users perform their transmissions without any restrictions on the starting time, and proposed a solution utilizing ODMA at the transmitter. The proposed scheme combines joint starting time and pattern detection, single-user decoding, and SIC to recover user messages, while performing double sliding window decoding. Simulation results show that our proposed scheme is superior to the alternative scheme in the literature for up to 100 active users per packet duration. 



\end{document}